\documentclass[11pt]{article}
\usepackage{moriond_olness,epsfig,multicol}

\def\lsim{\mathrel{\hbox{\rlap{\hbox{\lower4pt\hbox{$\sim$}}}\hbox{$<$}}}}
\def\gsim{\mathrel{\rlap{\lower4pt\hbox{\hskip1pt$\sim$}} \raise1pt\hbox{$>$}}}

\newcommand{\gevsq}{\ensuremath{\mathrm{GeV^2}}}
\def\eqref#1{Eq.~\ref{#1}}
\begin{document}
%
\hfill  CERN-PH-TH-2008-122
%
\vspace*{3cm}
\looseness=-1
\title{PARTON DISTRIBUTION FUNCTION UNCERTAINTIES \\ 
\& NUCLEAR CORRECTIONS FOR 
THE LHC${}^\dagger$\footnotetext{${}^\dagger$Presented by Fred Olness.}}
\author{
I.~SCHIENBEIN,$^{a,b}$ 
J.~Y.~YU,$^a$
C.~KEPPEL,$^{c,d}$
J.~G.~MORFIN,$^e$
F.~OLNESS,$^{a,g}$ 
and
J.F.~OWENS$^f$
}
\address{
$^a$Southern Methodist University, Dallas, TX 75206, USA,
\\
$^b$LPSC, Universit\'e Joseph Fourier
Grenoble, France
\\
$^c$Thomas Jefferson National Accelerator Facility, Newport News, VA 23602, USA,
\\
$^d$Hampton University, Hampton, VA, 23668, USA,
\\
$^e$Fermilab, Batavia, IL 60510, USA,
\\
$^f$Florida State University, Tallahassee, FL 32306-4350, USA
\\
$^g$Theoretical Physics Division, Physics Department, 
CERN, CH 1211 Geneva 23, Switzerland}
\maketitle\abstracts{
We study nuclear effects of charged current deep inelastic
neutrino-iron scattering in the framework of a $\chi^2$ analysis of
parton distribution functions (PDFs). We extract a set of iron PDFs which are
used to compute $x_{Bj}$-dependent and $Q^2$-dependent nuclear
correction factors for iron structure functions which are required in
global analyses of free nucleon PDFs.  We compare our results with
nuclear correction factors from neutrino-nucleus scattering models and
correction factors for $\ell^\pm$-iron scattering.  We find that,
except for very high $x_{Bj}$, our correction factors differ in both
shape and magnitude from the correction factors of the models and
charged-lepton scattering.
}


\footnotetext{\hangindent=10mm  Talk given at the {\it XLIIIth Rencontres de
Moriond QCD and High Energy Interactions}, \\ La Thuile, March 8-15,
2008. }

\footnotetext{\hangindent=10mm  Talk given at the {\it 16th International Workshop on
Deep-Inelastic Scattering (DIS2008)},  \\ 7-11 April 2008, University
College London.}

\section{Impact of Nuclear Corrections on PDFs}

\def\figone{
\begin{figure}[t]
\begin{center}
\includegraphics[width=0.46\textwidth]{./figs/e866_ref2TMC_2.eps}
\hfill
\includegraphics[width=0.48\textwidth]{./figs/weightave_tst_ref_all_xsecs2.eps}
\end{center}
\caption{
a) Comparison between the reference fit and the E-866 data. 
b) Comparison between the reference fit and the Chorus and NuTeV
neutrino cross section data.}
\label{fig:owens}
\end{figure}
}

\def\figtwo{
\begin{figure}[t]
\begin{minipage}[t]{0.49\linewidth}
\centering
\includegraphics[width=0.92\textwidth]{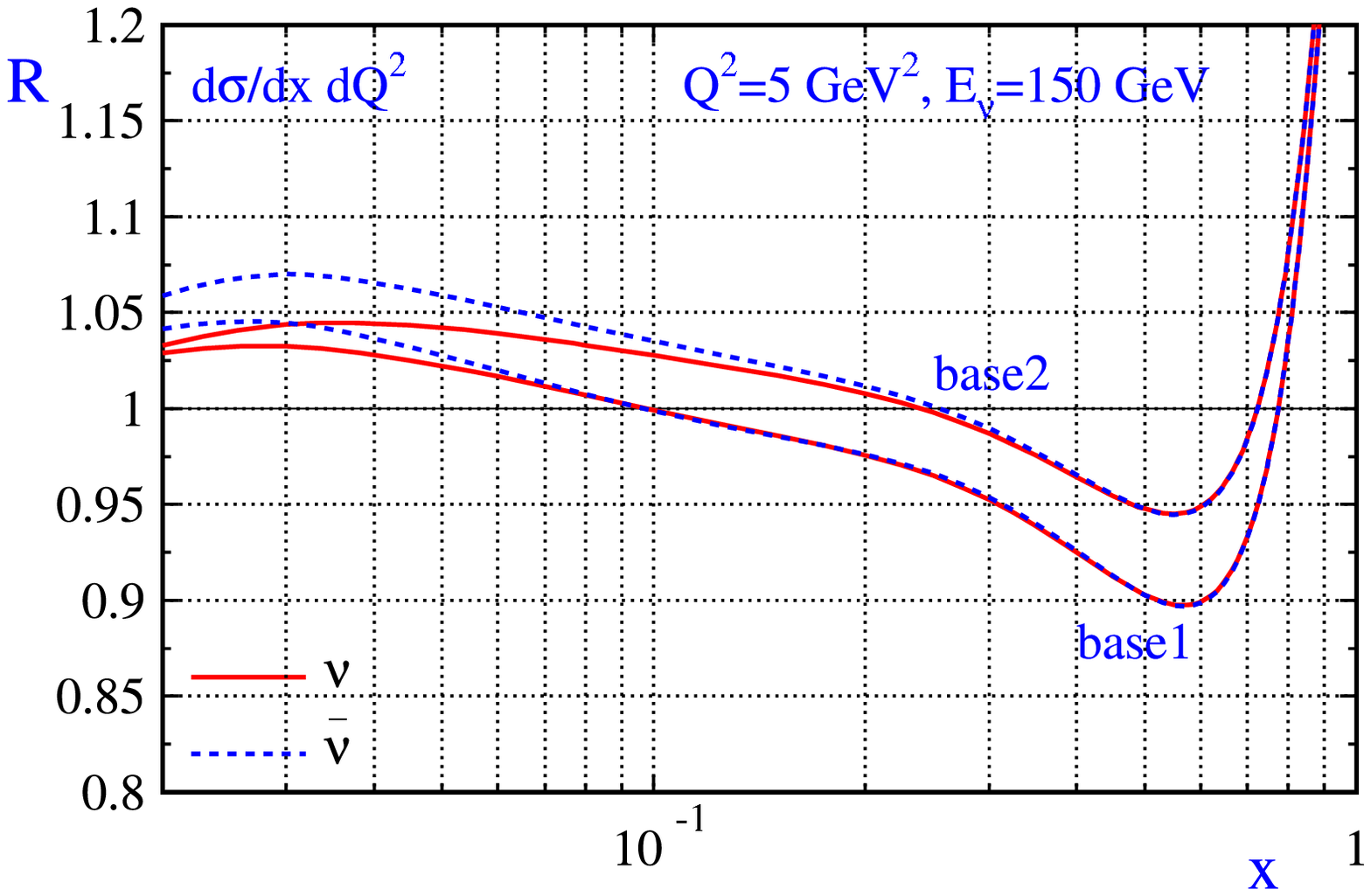}
\caption{
Nuclear correction factor $R$ according to \protect\eqref{eq:R}
for the differential cross section $d^2\sigma/dx \, dQ^2$ in charged
current neutrino-Fe scattering at $Q^2=5~\gevsq$.
 Results are shown for the charged current neutrino (solid lines) and
anti-neutrino (dashed lines) scattering from iron.
 The upper (lower) pair of curves shows the result of our analysis
with the Base-2 (Base-1) free-proton PDFs. 
\hfill \hfill
}
\label{fig:figure1}
\end{minipage}
\hspace{0.1cm}
\begin{minipage}[t]{0.49\linewidth}
\centering
\includegraphics[width=0.92\textwidth]{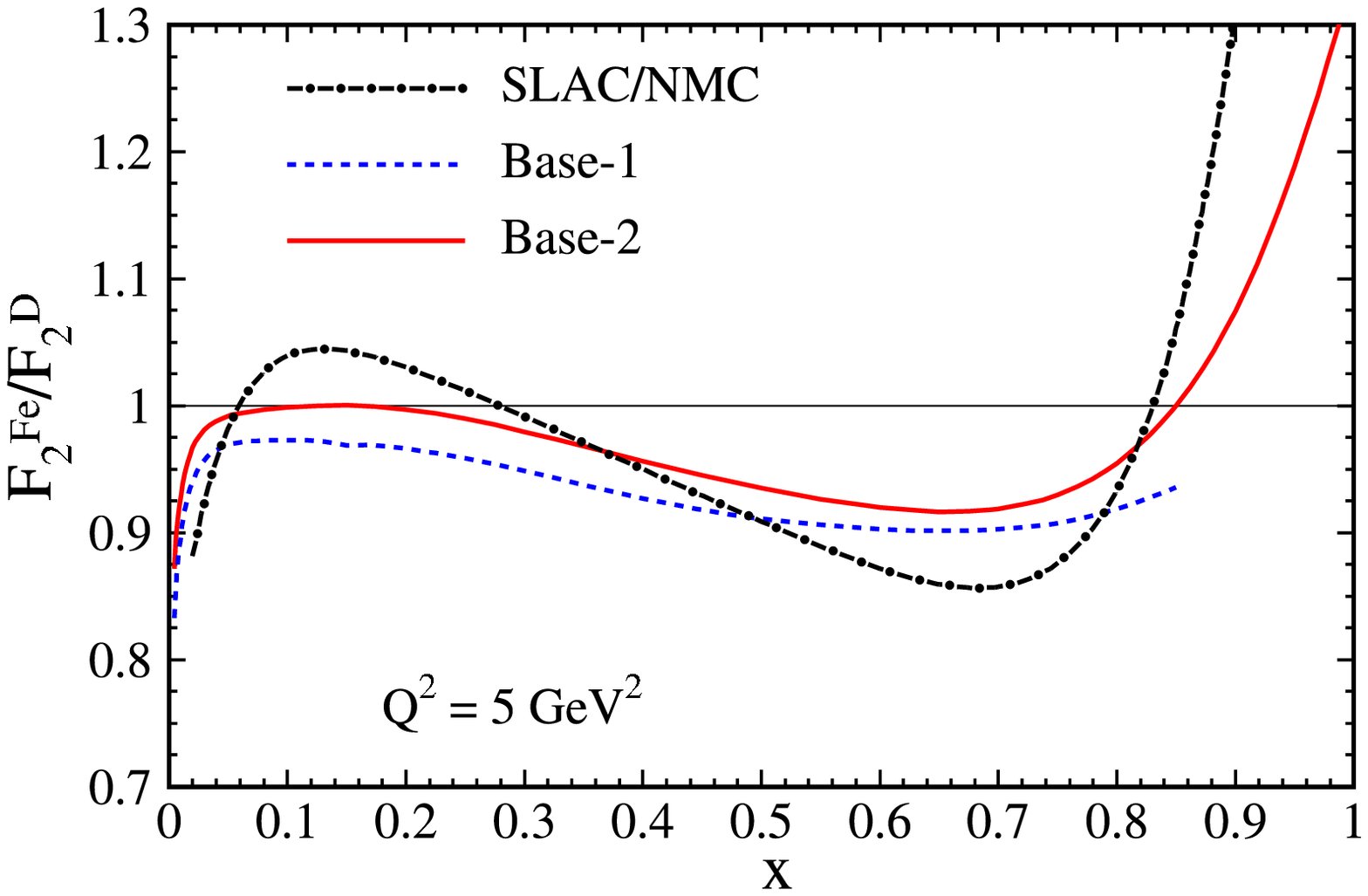}
\caption{
Predictions (solid and dashed line) for the  structure function ratio 
$F_2^{Fe}/F_2^{D}$ using the iron PDFs extracted from fits to NuTeV
neutrino and anti-neutrino data.
The SLAC/NMC parameterization is shown with the dot-dashed line. 
The structure function $F_2^D$ in the denominator has been computed
using either the Base-2 (solid line) or the Base-1 (dashed line) PDFs.
\hfill \hfill
}
\label{fig:figure2}
\end{minipage}
\end{figure}
} 

\def\figthree{
\begin{figure}[t]
\begin{center}
\includegraphics[width=0.49\textwidth]{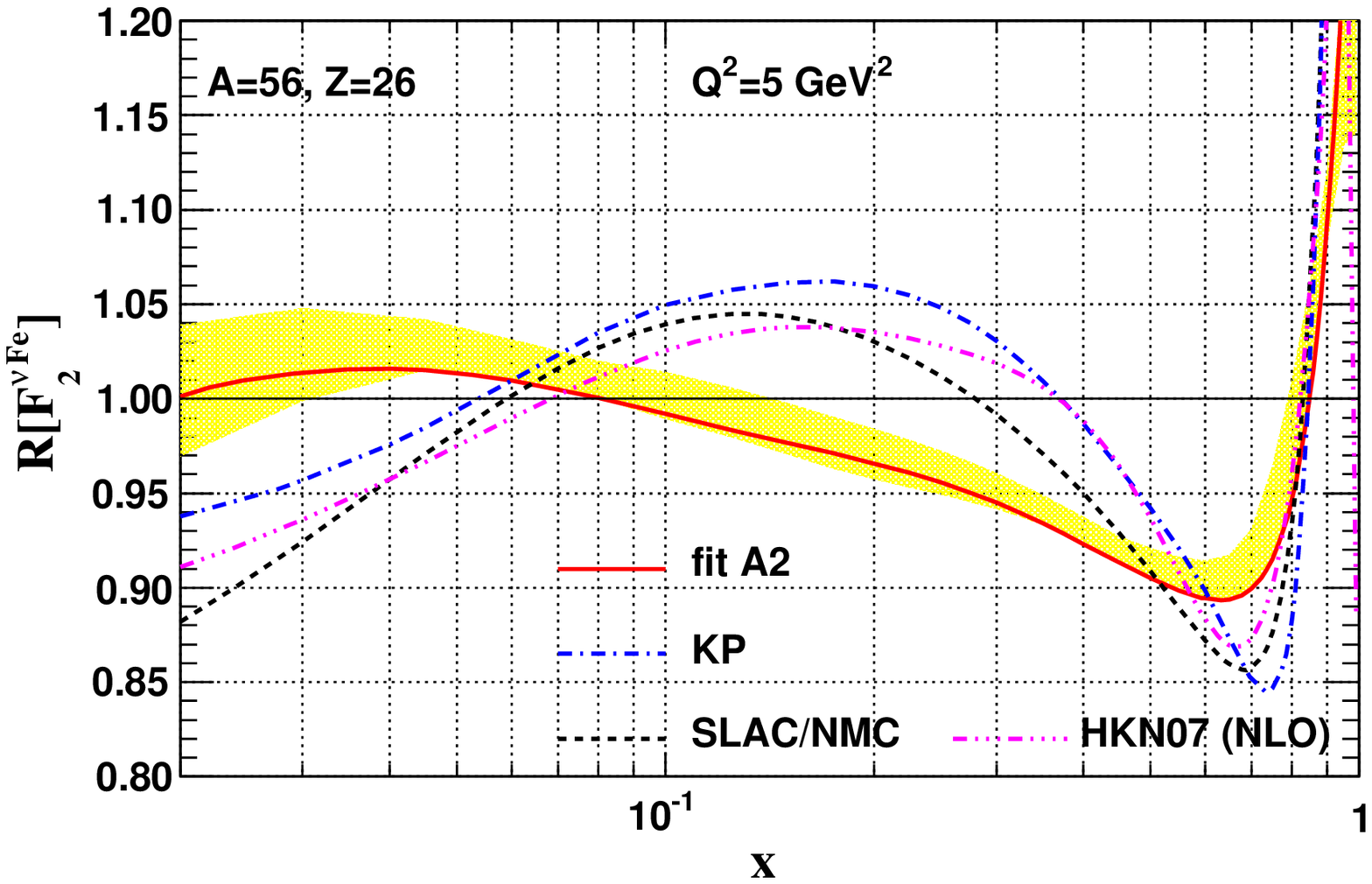}
\hfill
\includegraphics[width=0.49\textwidth]{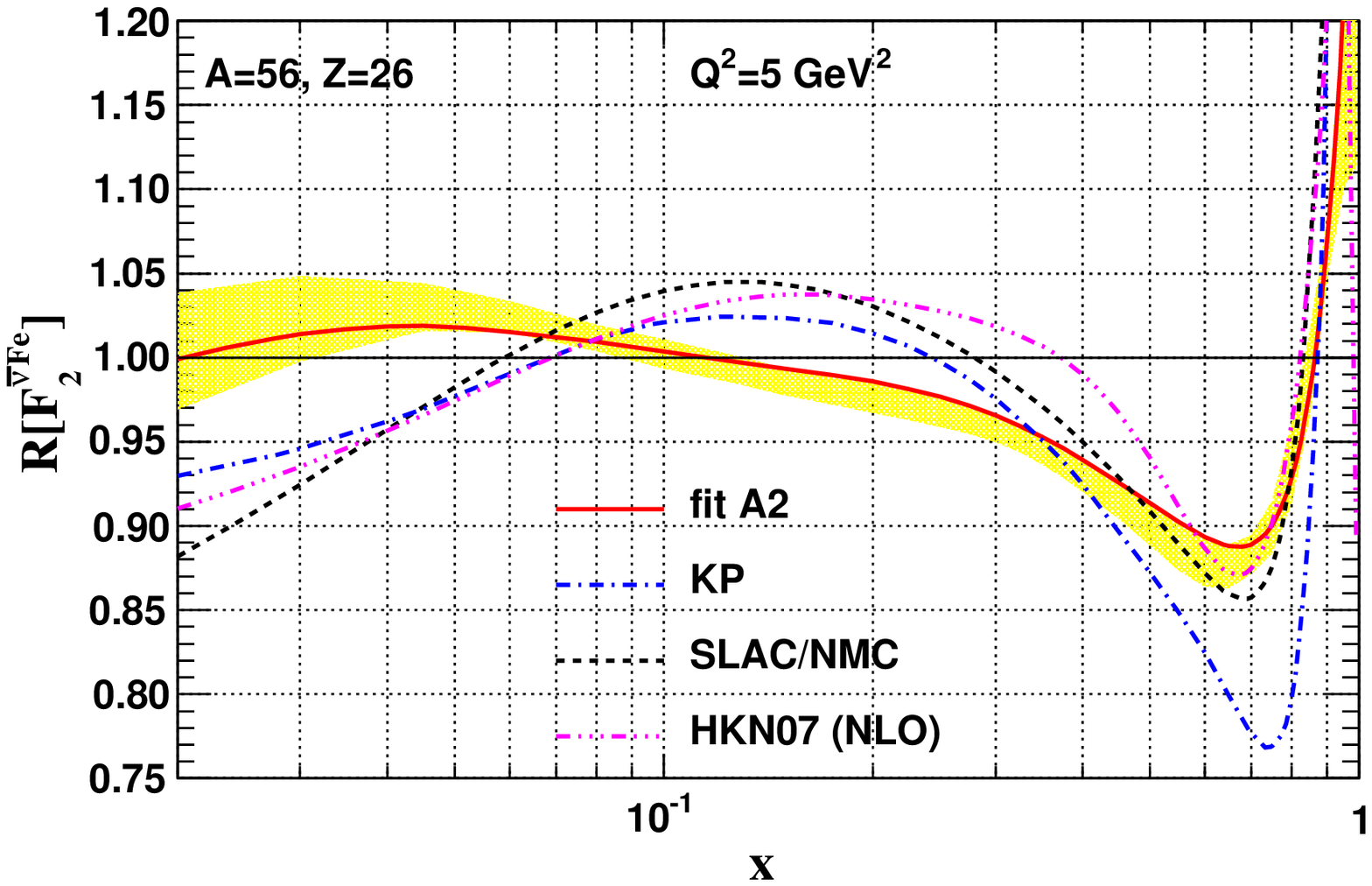}
\end{center}
\caption{
Nuclear correction factor $R$ for the structure function $F_2$ in
 a)~neutrino and b)~anti-neutrino  scattering from Fe.
The solid curve shows the result of our analysis of NuTeV data; the
uncertainty from the fit is represented by the shaded (yellow) band.  For
comparison we show the correction factor from the 
Kulagin--Petti model (dashed-dot line), \protect\cite{kp} 
HKN07 (dashed-dotted line),\protect\cite{kumano}
 and the SLAC/NMC parametrization (dashed line).\protect\cite{npdf}
}
\label{fig:figure3}
\end{figure}
} 

\def\figXXX{
\begin{figure}[ht]
\begin{minipage}[t]{0.5\linewidth}
\centering
\includegraphics[width=0.92\textwidth]{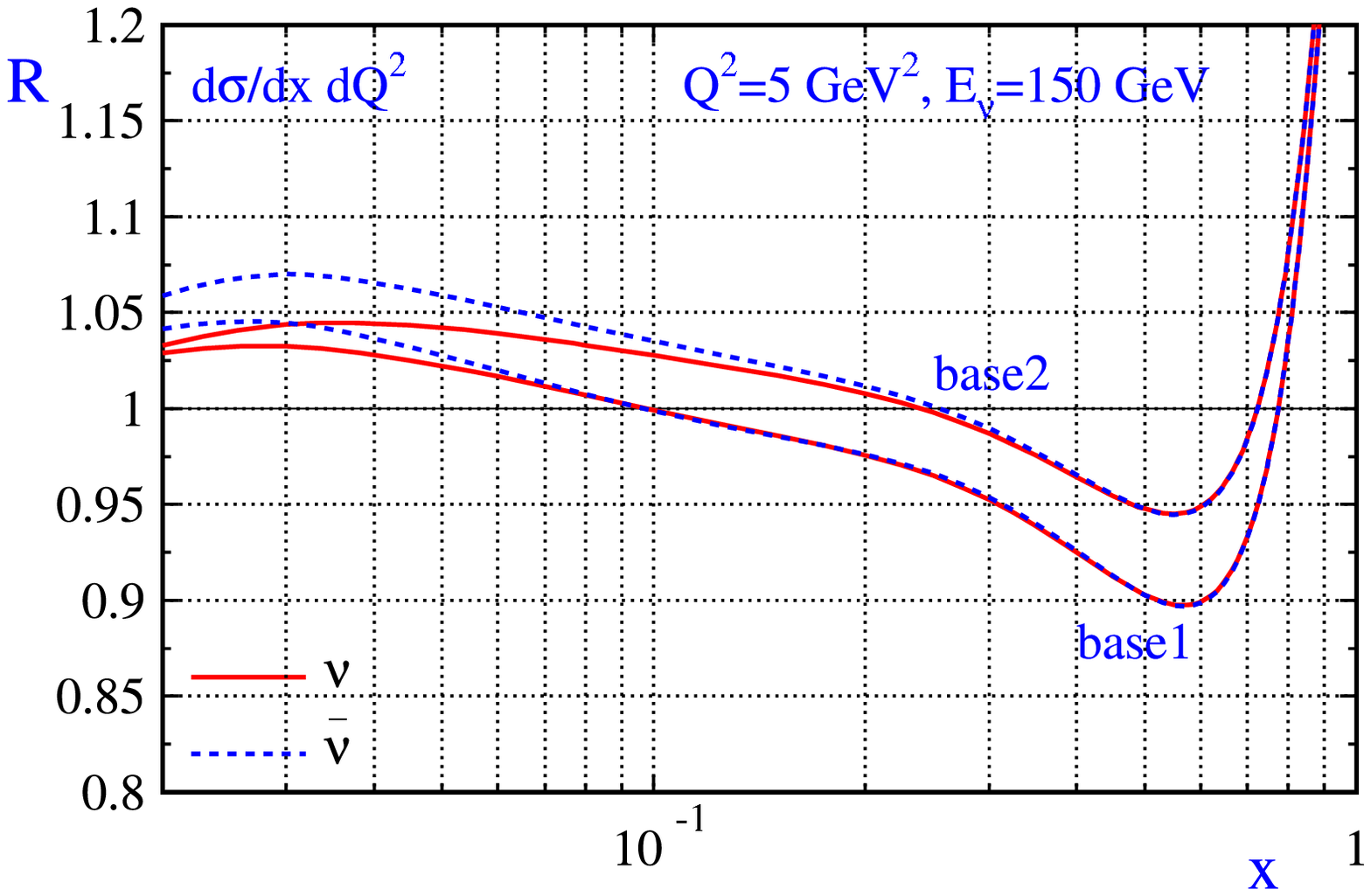}
\caption{
default
}
\label{fig:figure1}
\end{minipage}
\hspace{0.5cm}
\begin{minipage}[t]{0.5\linewidth}
\centering
\includegraphics[width=0.99\textwidth]{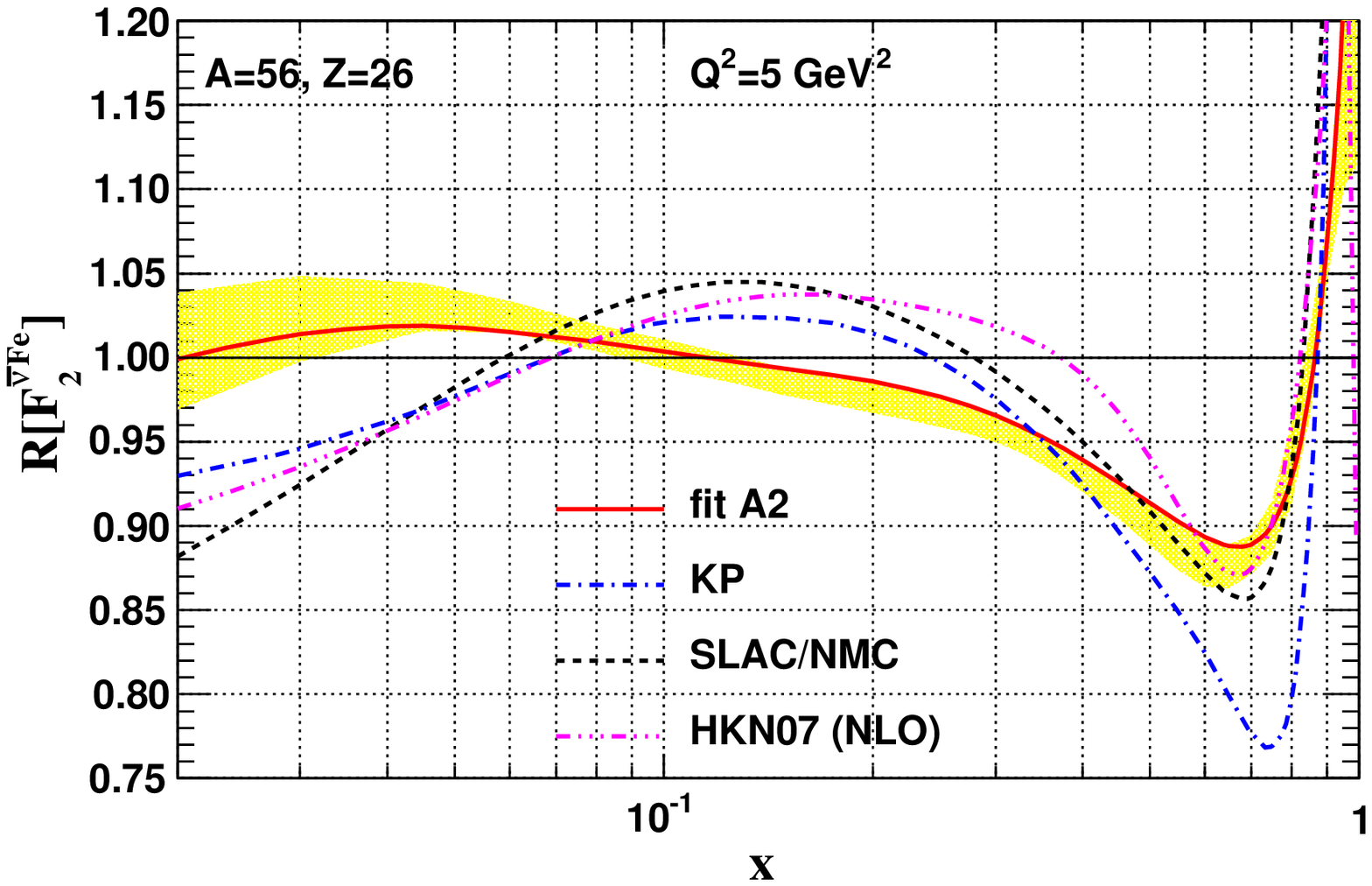}
\caption{
default
}
\label{fig:figure2}
\end{minipage}
\end{figure}
} 

%

\figtwo

The high statistics measurements of neutrino deeply inelastic
scattering (DIS) on heavy nuclear targets has generated significant
interest in the literature since these measurements provide valuable
information for global fits of parton distribution functions (PDFs).
The use of nuclear targets is unavoidable due to the weak nature of
the neutrino interactions, and this complicates the extraction of free
nucleon PDFs because model-dependent corrections must be applied to
the data.
In early PDF analyses, the nuclear corrections were static correction
factors without any (significant) dependence on the energy scale $Q$,
the atomic number $A$, or the specific observable.
The increasing precision of both the experimental data and the
extracted PDFs demand that the applied nuclear correction factors be
equally precise as these contributions play a crucial role in
determining the PDFs.
In this study we reexamine the source and size of the nuclear
corrections that enter the PDF global analysis, and quantify the
associated uncertainty. Additionally, we provide the foundation for
including the nuclear correction factors as a dynamic component of the
global analysis so that the full correlations between the heavy and
light target data can be exploited.

A recent study \cite{owens} analyzed the impact of new data sets from
the NuTeV~\cite{nutev}, Chorus, and E-866 Collaborations on the PDFs.
This study found that the NuTeV data set (together with the model used
for the nuclear corrections) pulled against several of the other data
sets, notably the E-866, BCDMS and NMC sets. Reducing the nuclear
corrections at large values of $x$ reduced the severity of this pull
and resulted in improved $\chi^2$ values.  These results suggest on a
purely phenomenological level that the appropriate nuclear corrections
for $\nu$-DIS may well be smaller than assumed.


To investigate this question further, we use the high-statistics
$\nu$-DIS experiments to perform a dedicated PDF fit to neutrino--iron
data.\cite{npdf}
Our methodology for this fit is parallel to that of the previous
global analysis,\cite{owens} {\it but} with the difference 
we use only Fe data and
that no
nuclear corrections are applied to the analyzed data; hence, the
resulting PDFs are for a bound proton in an iron nucleus.
Specifically, we determine iron PDFs using the recent NuTeV
differential neutrino (1371 data points) and anti-neutrino (1146 data
points) DIS cross section data,\cite{nutev} and we include NuTeV/CCFR
dimuon data (174 points) which are sensitive
to the strange quark content of the nucleon.
We impose kinematic cuts of $Q^2>2$~GeV and $W>3.5$~GeV, and  
obtain a good fit with a $\chi^2$ of 1.35 per data point.\cite{npdf}

\section{Nuclear Correction Factors \label{sec:nfac}}

\figthree

We now compare our iron PDFs with the free-proton PDFs (appropriately
scaled) to infer the proper heavy target correction which should be
applied to relate these quantities.
Within the parton model, a nuclear correction factor $R[{\cal O}]$ for
an observable ${\cal O}$ can be defined as follows:
\begin{equation}
R[{\cal O}] = \frac{{\cal O}[{\rm NPDF}]}{{\cal O}[{\rm free}]}
\label{eq:R}
\end{equation}
where ${\cal O}[{\rm NPDF}]$ represents the observable computed with
nuclear PDFs, and ${\cal O}[{\rm free}]$ is the same observable
constructed out of the free nucleon PDFs.
In addition to the kinematic variables and the factorization scale,
$R$ can depend on the observable under consideration simply because
different observables may be sensitive to different combinations of
PDFs.
This means that the nuclear correction factor $R$ for $F_2^A$ and
$F_3^A$ will, in general, be different.  Additionally, the nuclear
correction factor for $F_2^A$ will yield different results for the
charged current $\nu$--$Fe$ process ($W^\pm$ exchange) as compared
with the neutral current $\ell^\pm$--$Fe$ process ($\gamma$ exchange).
Because we have extracted the iron PDFs from only iron data, we do
not assume any particular form for the nuclear $A$-dependence; hence
the extracted $R[{\cal O}]$ ratio is essentially model independent.

We begin by computing the nuclear correction factor $R$ for the
neutrino differential cross section $d^2\sigma/dx \, dQ^2$, as this
represents the bulk of the NuTeV data included in our fit, 
{\it cf.,}  Fig.~\ref{fig:figure1}.
We have computed $R$ using two separate proton PDFs denoted as Base-1
and Base-2; the difference of these curves, in part, reflects the
uncertainty introduced by the proton 
PDF.\cite{npdf}
We also observe that the neutrino and anti-neutrino results coincide in
the region of large $x$ where the valence PDFs are dominant, but
differ by a few percent at small $x$ due to the differing strange and
charm distributions.

We next display the nuclear correction factors for 
$F_2^{\nu Fe}$ and 
$F_2^{\bar\nu Fe}$ in Fig.~\ref{fig:figure3}. 
The SLAC/NMC curve 
has been obtained from an $A$ and $Q^2$-independent parameterization 
of calcium and iron charged--lepton DIS data.\cite{owens}
Due to the neutron excess in iron,
both our curves and the KP curves differ when comparing scattering for
neutrinos and anti-neutrinos.
For our results (solid lines), the difference between the neutrino and
anti-neutrino results is relatively small, of order $3 \%$ at $x=0.6$.
Conversely, for the KP model (dashed-dotted lines) the
$\nu$--$\bar\nu$ difference reaches $10 \%$ at $x\sim 0.7$, and
remains sizable at lower values of $x$.

Comparing the nuclear correction factors for the $F_2$ structure
function (Fig.~\ref{fig:figure3}) with those
obtained for the differential cross section (Fig.~\ref{fig:figure1}), we
see these are quite different, particularly at small $x$. This
is because the cross section $d^2\sigma$ is comprised of a
different combination of PDFs than the $F_2$ structure function. 
Again, we emphasize that it is important to use an appropriate nuclear
correction factor which is matched to the particular observable.


Our results have general
features in common with the KP model and the SLAC/NMC
parameterization, but the magnitude of the effects and the $x$-region
where they apply are quite different.  Our results are noticeably
flatter than the KP and SLAC/NMC curves, especially at moderate-$x$
where the differences are significant.
The general trend we see when examining these nuclear correction
factors is that the anti-shadowing region is shifted to smaller $x$
values, and any turn-over at low $x$ is minimal given the PDF
uncertainties.
In general, these plots suggest that the size of the nuclear
corrections extracted from the NuTeV data are smaller than those
obtained from charged lepton scattering (SLAC/NMC) or from the set of
data used in the KP model.

Since the SLAC/NMC parameterization was fit to $F_2^{Fe}/F_2^D$ for
charged-lepton DIS data, we can perform a more balanced comparison by
using our iron PDFs to compute this same quantity. The results are
shown in Fig.~\ref{fig:figure2} where we have used our iron PDFs to
compute $F_2^{Fe}$, and the Base-1 and Base-2 PDFs to compute
$F_2^{D}$.
As before,
we find our results have some gross features in common while on a more
refined level the magnitude of the nuclear corrections extracted from
the charged current iron data differs from the charged lepton data.
In particular, we note that the so-called ``anti-shadowing''
enhancement at $x \sim [0.06-0.3]$ is {\it not} reproduced by the
charged current (anti-)neutrino data.
Examining our results among all the various $R[{\cal O}]$
calculations, we generally find that any nuclear enhancement in the
small $x$ region is reduced and shifted to a lower $x$ range as
compared with the SLAC/NMC parameterization.
Furthermore, in the limit of large $x$ ($x \gsim 0.6$) our results are
slightly higher than the data, including the very precise SLAC-E139
points; however, the large theoretical uncertainties on $F_2^D$ in this
$x$-region  make it difficult to extract firm
conclusions.

\section{Conclusions}\label{sec:summary}

%
While the nuclear corrections extracted from charged current
$\nu$--$Fe$ scattering have similar general 
characteristics as the neutral
current $l^\pm$--$Fe$ charged-lepton results, the detailed $x$ and $Q^2$
behavior is quite different.
 There
is {\it a priori} no requirement that these be equal; in fact, given
that the $\nu$--$Fe$ process involves the exchange of a $W$ and the
$\ell^\pm$--$Fe$ process involves the exchange of a $\gamma$ we
necessarily expect this will lead to differences at some level.

These results raise the deeper question as to whether the charged
current and neutral current correction factors may be substantially
different.
A combined analysis of neutrino and charged-lepton data sets, for
which the present study provides a foundation, will shed more light on
these issues.
Resolving these questions is essential if we are to reliably use the
plethora of nuclear data to obtaining free-proton PDFs 
which form the basis of the LHC analyses.



We thank 
Tim Bolton,
Javier Gomez, 
Shunzo Kumano,
Eric Laenen,
Dave Mason, 
W. Melnitchouk, 
Donna Naples,
Mary Hall Reno,
Voica~A.~Radescu,
and
Martin Tzanov
for valuable discussions, 
and 
%
BNL, CERN, and Fermilab
for their hospitality.
This work was  supported by  U.S.\ DoE  
DE-FG02-04ER41299,   DE-FG02-97IR41022, 
DE-AC05-06OR23177, 
NSF 
grant 0400332,
Lightner-Sams Foundation, 
\&
Deutsche Forschungsgemeinschaft 
(YU~118/1-1).


\end{document}